\documentclass[aps,pre,twocolumn,superscriptaddress]{revtex4}

\usepackage[dvips]{graphicx}

\usepackage{pnastwof}

\usepackage{amssymb,amsfonts,amsmath}
\usepackage{color}

\DeclareMathOperator{\erf}{erf}
\DeclareMathOperator{\erfc}{erfc}
\newcommand{\vect}[1]{\mathbf{#1}}
\newcommand{\VR}{\mathcal{V}_R}

\newcommand{\laplacian}{\vec{\nabla}^2}

\begin{document}








\title{Interplay of local hydrogen-bonding and long-ranged dipolar forces in simulations of confined water}




\author{Jocelyn M. Rodgers}
\affiliation{Institute for Physical Science and Technology, University of Maryland, College Park, MD 20742}
\affiliation{Chemical Physics Program, University of Maryland, College Park, MD 20742}

\author{John D. Weeks}
\affiliation{Institute for Physical Science and Technology, University of Maryland, College Park, MD 20742}
\affiliation{Department of Chemistry and Biochemistry, University of Maryland, College Park, MD 20742}





\begin{abstract}
  Spherical truncations of Coulomb interactions in standard models for
  water permit efficient molecular simulations and can give remarkably
  accurate results for the structure of the uniform liquid. However
  truncations are known to produce significant errors in nonuniform
  systems, particularly for electrostatic properties.  Local molecular
  field (LMF) theory corrects such truncations by use of an effective
  or restructured electrostatic potential that accounts for effects of the
  remaining long-ranged interactions through a density-weighted
  mean field average and satisfies a modified Poisson's equation defined
  with a Gaussian-smoothed charge density.  We apply LMF theory to three simple molecular
  systems that exhibit different aspects of the failure of a naive
  application of spherical truncations -- water confined between
  hydrophobic walls, water confined between atomically-corrugated
  hydrophilic walls, and water confined between hydrophobic walls with
  an applied electric field.  Spherical truncations of $1/r$ fail
  spectacularly for the final system in particular, and LMF theory
  corrects the failings for all three.  Further, LMF theory provides a
  more intuitive way to understand the balance between local hydrogen
  bonding and longer-ranged electrostatics in molecular simulations
  involving water.
\end{abstract}

\maketitle


\section{Introduction}
An accurate and efficient treatment of Coulomb interactions in
molecular simulations represents an important and ongoing challenge.
Standard biomolecular simulation packages like {\sc
  charmm}~\cite{MacKerellBashfordBellott.1998.All-Atom-Empirical-Potential-for-Molecular-Modeling}
and {\sc
  amber}~\cite{DuanWuChowdhury.2003.A-point-charge-force-field-for-molecular-mechanics}
quite generally assign effective point charges to interaction sites
even in neutral molecules in order to approximate the charge
separation of the electron cloud along polar bonds.  Effective point
charges are also found in most standard water models such as the
extended simple point charge (SPC/E)
model~\cite{BerendsenGrigeraStraatsma.1987.The-missing-term-in-effective-pair-potentials}
shown in Fig.~1\emph{a}, and water molecules are increasingly being
included explicitly in biomolecular simulations.

\begin{figure}[bth]
  \centering
  \includegraphics{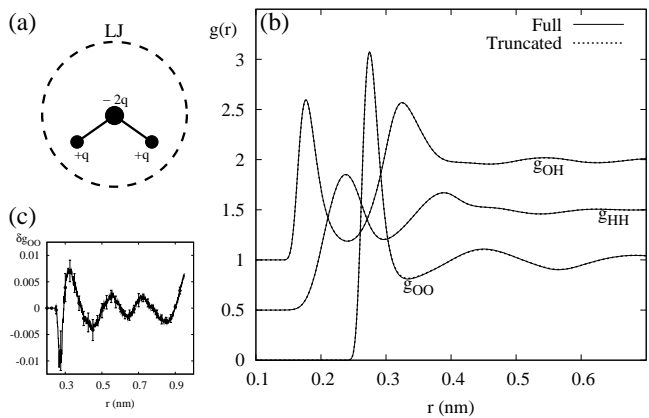}
  \caption{Bulk SPC/E water at 300 K and 0.998 g/cm$^3$. (a)~Diagram of SPC/E model of water. 
    Point charges are assigned to each atomic site and a Lennard-Jones
    potential is centered about the oxygen with core diameter
    $\sigma_{\text{LJ}}$ indicated by dashed lines.  (b)~Various site-site
    pair correlation functions in bulk SPC/E water are each displaced
    vertically by 0.5 for clarity.  Gaussian-truncated (GT) water
    using $v_0(r)$ with $\sigma=0.40$~nm agrees well with a full
    treatment of electrostatics. (c)~Deviations $\delta
    \text{g}_{\text{OO}}$, defined as the difference between g(r) for
    GT water and g(r) using full electrostatics.  Simulation error
    bars are also shown.}
\end{figure}

In order to minimize edge effects in necessarily finite simulation
cells, these simulations use periodic boundary
conditions~\cite{FrenkelSmit.2002.Understanding-Molecular-Simulation:-From-Algorithms}.
Researchers have long sought to devise spherical truncations of the
Coulomb $1/r$ potential so that the truncated potential accurately
describes the strong Coulomb forces between closely-spaced charges
that lead to hydrogen bonding in water but then vanishes quickly
beyond some properly chosen cutoff radius.  Then, only the minimum
(closest) image of an interacting charge needs to be accounted for
and fast and efficient simulations that scale linearly with system size are possible.
In bulk-like systems these spherical truncation approaches can be surprisingly accurate, but
standard estimates of thermodynamic properties in the bulk
liquid are less satisfactory~\cite{ZahnSchillingKast.2002.Enhancement-of-the-wolf-damped-Coulomb-potential:,FennellGezelter.2006.Is-the-Ewald-summation-still-necessary-Pairwise,WolfKeblinskiPhillpot.1999.Exact-method-for-the-simulation-of-Coulombic-systems,Nezbeda.2005.Towards-a-Unified-View-of-Fluids,HummerPrattGarcia.1996.Free-energy-of-ionic-hydration,HummerSoumpasisNeumann.1994.Computer-simulation-of-aqueous-Na-Cl-Electrolytes,TironiSperbSmith.1995.A-generalized-reaction-field-method-for-molecular-dynamics,IzvekovSwansonVoth.2008.Coarse-graining-in-interaction-space:-A-systematic-approach}.

However, it has long been
established~\cite{FellerPastorRojnuckari.1996.Effect-of-Electrostatic-Force-Truncation-on-Interfacial}
that when these spherical truncations are employed in nonuniform
geometries, such as near a lipid bilayer, there can be pronounced
errors in structure, thermodynamics, and particularly electrostatic
properties. In those cases, the neglected long-ranged
forces~\cite{Spohr.1997.Effect-of-Electrostatic-Boundary-Conditions-and-System}
compete with the local ordering due to cores and hydrogen bonding
captured well by spherical truncations.

These failures have led most researchers to conclude that the success
of spherical truncations in bulk was seductive but misleading.
Instead, lattice summation techniques -- the electrostatic method of choice --
give an accurate account of all periodic images.
Ewald~\cite{Ewald.1921.Evaluation-of-optical-and-electrostatic-lattice-potentials}
and
Lekner~\cite{Lekner.1991.Summation-of-Coulomb-fields-in-computer-simulated-disordered-systems}
summations have long been used in simulations of charged and dipolar
fluids and biomolecules.  Much current research is focused on
developing more efficient implementations such as particle-mesh
Ewald~\cite{DardenYorkPedersen.1993.Particle-mesh-Ewald:-An-Ndot-log-N-method,DesernoHolm.1998.How-to-mesh-up-Ewald-sums.-I.-A-theoretical-and-numerical}
and in removing some of the artifacts that may arise from the use of
periodic lattice sums in fluid
environments~\cite{WuBrooks.2005.Isotropic-Periodic-Sum:-A-Method-for-the-Calculation}.

But is it possible to build on the success of spherical truncations of
$1/r$ interactions in uniform
environments~\cite{ZahnSchillingKast.2002.Enhancement-of-the-wolf-damped-Coulomb-potential:,FennellGezelter.2006.Is-the-Ewald-summation-still-necessary-Pairwise,WolfKeblinskiPhillpot.1999.Exact-method-for-the-simulation-of-Coulombic-systems,Nezbeda.2005.Towards-a-Unified-View-of-Fluids,HummerPrattGarcia.1996.Free-energy-of-ionic-hydration,HummerSoumpasisNeumann.1994.Computer-simulation-of-aqueous-Na-Cl-Electrolytes,TironiSperbSmith.1995.A-generalized-reaction-field-method-for-molecular-dynamics,IzvekovSwansonVoth.2008.Coarse-graining-in-interaction-space:-A-systematic-approach}
and to understand and correct their failures in nonuniform
environments~\cite{FellerPastorRojnuckari.1996.Effect-of-Electrostatic-Force-Truncation-on-Interfacial,Spohr.1997.Effect-of-Electrostatic-Boundary-Conditions-and-System}?
We show that local molecular field (LMF)
theory~\cite{WeeksKatsovVollmayr.1998.Roles-of-repulsive-and-attractive-forces-in-determining,
  WeeksSelingerBroughton.1995.Self-Consistent-Treatment-of-Repulsive-and-Attractive-Forces,Weeks.2002.Connecting-local-structure-to-interface-formation:,ChenKaurWeeks.2004.Connecting-systems-with-short-and-long}
offers exactly that opportunity. LMF theory corrects naive
truncation schemes by introducing an effective electrostatic potential
$\VR(\vect{r})$ that accounts for the remaining effects of the
long-ranged interactions.  LMF theory also suggests a particularly
effective smoothed Gaussian-truncation scheme for the charges in SPC/E
water and other interaction-site molecular models.  Furthermore, the
theory provides a general conceptual framework that permits a
qualitative as well as quantitative understanding of the effects of
long-ranged Coulomb interactions on local water structure -- something
that is often obscured in black box application of lattice summation
techniques.

We will briefly discuss the application of the Gaussian-truncation
scheme to bulk SPC/E water. Then we will demonstrate that the
inclusion of the effective potential $\VR(\vect{r})$ given by LMF
theory is the sole modification of molecular dynamics simulations of
Gaussian-truncated (GT) water required to yield accurate minimum image
simulations of water in nonuniform slab environments.  While we focus
on systems composed of solely water molecules and simple confinement
potentials, we argue below that the LMF approach is relevant for the
treatment of electrostatics in general biomolecular simulations.

\section{Bulk Gaussian-Truncated (GT) Water}
In LMF theory, we split the $1/r$ potential associated with each
charge into rapidly-varying ($v_0$) and slowly-varying ($v_1$)
components as
\begin{equation}
  \label{eqn:potslit}
  \frac{1}{r} = \frac{\erfc(r/\sigma)}{r} + \frac{\erf(r/\sigma)}{r} = v_0(r) + v_1(r).
\end{equation}
This potential separation isolates strong short-ranged and rapidly-varying
interactions in $v_0(r)$, and the remaining slowly-varying long-ranged
forces are encompassed by $v_1(r)$.  This interaction $v_1(r)$ is proportional
to the electrostatic potential arising from a smooth normalized Gaussian charge distribution with
width $\sigma$, and is defined via the convolution
\begin{equation}
  \label{eqn:v1def}
 v_1(r) \equiv  \frac{1}{\pi^{3/2}\sigma^3}\int e^{-r^{\prime 2}/\sigma^2}\frac{1}{\left| \vect{r} - \vect{r^\prime} \right|} \, d\vect{r}^\prime.
\end{equation}
Thus  $v_{1}(r)$ is slowly-varying in $r$-space over
the \emph{smoothing length} $\sigma$ and is simultaneously localized to
small wave vectors in reciprocal space, as can be seen from its Fourier transform
$\hat{v}_{1}(k) = 4\pi {k^{-2}} \exp [{-(k\sigma )^{2}/4}]$.

Hence $v_0(r) \equiv 1/r - v_1(r)$ is the screened potential resulting
from a point charge surrounded by a neutralizing Gaussian charge
distribution whose width $\sigma$ also sets the scale for the smooth truncation of $v_{0}$.
For small $r < \sigma$, the force due to $v_0(r)$
approaches that of the bare point charge.  By increasing $\sigma$ we
increase the effective range of essentially unscreened Coulomb
interactions included in $v_0(r)$.

LMF theory introduces a general mapping from the original system with
full Coulomb interactions and an applied electrostatic potential
$\mathcal{V}(r)$ arising from fixed charges or an applied electric
field to a well-chosen ``mimic system'' with Coulomb interactions
replaced by the short-ranged $v_0(r)$:
\begin{equation}
  \label{eqn:mapping}
  \left\{\begin{array}{c}
      1/r  \\
      \\
      \mathcal{V}(\vect{r})
    \end{array}\right\}
  \xrightarrow{\text{LMF}}
  \left\{\begin{array}{c}
      v_0(r)  \\
      \\
      \VR(\vect{r})
    \end{array}\right\}.
\end{equation}
Here, $\VR(\vect{r})$ is a restructured electrostatic potential in the
mimic system that accounts for the averaged effects of the neglected
long-ranged forces represented by $v_1(r)$, in essence by solving a
modified Poisson's equation as explained in a later section.

In this paper, we simulate the SPC/E model of water as shown in
Fig.~1\emph{a}.  All data are results from molecular dynamics
simulations as described in the Methods section.  The point charges in
SPC/E water play a dual role, generating both strong short-ranged
forces that lead to the local hydrogen-bond network as well as the
longer-ranged dipole-dipole interactions.  Neither the point charges
nor the Lennard-Jones (LJ) potential describing the molecular cores
are perfect representations of actual water interactions but adjusting
the constants makes the total potential quite reasonable for
temperatures and densities of interest.

Since the charges generating the local hydrogen bonding network in
water are very strongly coupled, a reasonable first approximation is
simply to replace all point-charge interactions by the truncated
$v_0(r)$ without employing a $\VR$. LMF theory suggests that Coulomb
interactions are best truncated on a site-site basis, since it is the
basic $1/r$ potential that we attenuate about each site.  Such a
site-site scheme has been implemented in other spherical truncations
of $1/r$ interactions as
well~\cite{HummerSoumpasisNeumann.1994.Computer-simulation-of-aqueous-Na-Cl-Electrolytes}.
When the $v_0(r)$ truncations are made, we have Gaussian-truncated
(GT) water.

Crucial to the success of the Gaussian-truncation scheme is the choice
of a $\sigma$ that will be large enough to accurately describe the
nearest-neighbor hydrogen bonds.  Using a $\sigma$ of 0.4~nm or
larger, GT water can give a remarkably accurate description of all O
and H pair correlation functions in bulk water, as shown in
Fig.~1\emph{b}.  The difference between $\text{g}_{\text{OO}}$ from a full
electrostatic treatment (indicated by the labels ``Full'' in the
figures) and $\text{g}_{\text{OO}}$ from GT water is shown in Fig.~1\emph{c}
to emphasize the good agreement. 
We expect that certain earlier truncation
schemes~\cite{HummerSoumpasisNeumann.1994.Computer-simulation-of-aqueous-Na-Cl-Electrolytes}
with very slowly-varying $v_{1}(r)$ could give comparably accurate results.

The local hydrogen-bond network is often viewed as the most important
qualitative feature of
water~\cite{Ball.2008.Water-as-an-active-constituent-in-cell-biology},
and GT water captures that feature with a very good description of
nearest neighbor hydrogen-bond energetics and local tetrahedral order,
while ignoring the longer-ranged electrostatic effects of the bound
charges.  As argued elsewhere~\cite{HuRodgersWeeks..}, the accuracy of
GT water in bulk arises from a strong cancellation of the long-ranged
electrostatic forces exerted by surrounding molecules in the uniform
bulk on any given molecular site.  These site-based spherical
truncations can also yield highly accurate angular correlation
functions~\cite{HuRodgersWeeks..}, a property that is missed with
molecule-based truncations that do not neglect only slowly-varying
forces~\cite{Nezbeda.2005.Towards-a-Unified-View-of-Fluids}.  LMF
theory also suggests simple analytic corrections to the energy and
pressure such that the thermodynamics of bulk GT water are accurate as
well~\cite{RodgersWeeks..}.

\section{LMF-Corrected GT Water in an Applied Field}
Following previous
work~\cite{LeeMcCammonRossky.1984.The-Structure-of-Liquid-Water-at-an-Extended-Hydrophobic,YehBerkowitz.1999.Dielectric-constant-of-water-at-high-electric},
we confine SPC/E water between smoothed hydrophobic LJ
walls and apply a constant electric field $E_0$ normal to the walls,
as sketched in Fig~2\emph{a}.  The confined water should behave like a
dielectric slab from introductory electrostatics, resulting in a
polarization field $E_{\text{pol}}$ opposing the applied electric
field~\cite{CorsonLorrain.1962.Introduction-to-Electromagnetic-Fields-and-Waves}.
The oxygen density profile and charge density profile using only
GT water are completely incorrect when we apply an electric
field of 10.0 V/nm as shown in Fig.~2\emph{b} and Fig.~2\emph{c}.

\begin{figure}[tb]
  \centering
  \includegraphics{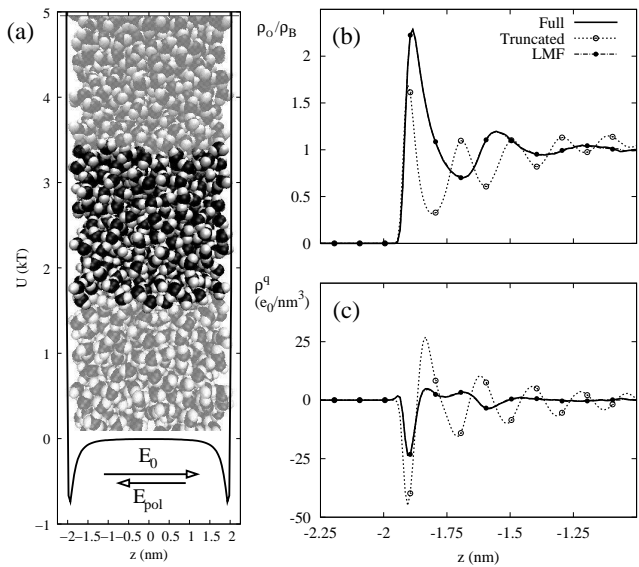}
  \caption{Model hydrophobic wall confinement of SPC/E water with an applied field.  GT water with $\sigma=0.6$~nm fails while treatment with full LMF theory succeeds.  (a)~The simulation system and nearest images in the $y$-direction confined by the Lennard-Jones wall potential.  (b)~The oxygen density profile relative to the bulk density.  (c)~The charge density profile.  Density profiles are only shown for $z < -1.0$~nm for greater clarity of the atomic level behavior, though these densities are strictly asymmetric about the origin.}
\end{figure}

Using spherical truncations alone neglects the long-ranged
electrostatic consequences of the local ordering of the water
molecules, which act to attenuate and weaken the applied electric
field in the central region as suggested by Le Chatelier's principle.
Thus the GT water molecules spectacularly over-order with the
unattenuated electric field since the externally applied field induces
an incorrect, strongly polarized state throughout the GT water slab that
happens to be compatible with an energetically-favored arrangement of
hydrogen bonding.  Inclusion of a self-consistent $\VR(z)$ to
represent the long-ranged net effect of water molecule polarization
makes the density profiles again agree very well with the results for
the full system as shown in Fig.~2\emph{b} and Fig.~2\emph{c}.  With
the inclusion of net long-ranged effects, the highly-ordered state
does not occur because it is entropically disfavored and thus is
highly unlikely in the true attenuated field.

The inclusion of a restructured $\VR(z)$ also corrects many other
properties of confined water when an electric field is applied such as
the dielectric constants and details of molecular orientation
profiles~\cite{RodgersWeeks..}.  However, the density profile is the
simplest and most spectacular example of the failure of GT water in
this system, and LMF theory handily corrects that failure.

\section{Local Molecular Field Theory for Electrostatics}
LMF theory has been applied to many different ionic systems, including
charge
mixtures~\cite{ChenKaurWeeks.2004.Connecting-systems-with-short-and-long,ChenWeeks.2006.Local-molecular-field-theory-for-effective,RodgersKaurChen.2006.Attraction-between-like-charged-walls:-Short-ranged,DenesyukWeeks.2008.A-new-approach-for-efficient-simulation-of-Coulomb-interactions}.
General derivations of the approach are available in the
literature~\cite{WeeksKatsovVollmayr.1998.Roles-of-repulsive-and-attractive-forces-in-determining,WeeksSelingerBroughton.1995.Self-Consistent-Treatment-of-Repulsive-and-Attractive-Forces,Weeks.2002.Connecting-local-structure-to-interface-formation:}.
However, when the LMF truncation is applied to the general $1/r$
functional form and the same $\sigma$ is applied to \emph{every}
charge-charge interaction in the system regardless of identity, we may
write a substantially simpler formulation of the LMF mixture equation
than previously
stated~\cite{ChenWeeks.2006.Local-molecular-field-theory-for-effective}.
An explanation of these symmetries and the validity of the LMF
equation below for both charge mixtures and site-site molecular models
will be detailed in forthcoming
papers~\cite{HuRodgersWeeks..,RodgersWeeks.2008.Local-molecular-field-theory-for-the-treatment}.
  
Provided that the $\sigma$ chosen is at least large enough for
$v_0(r)$ to incorporate the nearest-neighbor interactions such as
hydrogen-bonding in water and for $v_1(r)$ to be slowly varying over
those nearby correlations, LMF theory predicts that the restructured potential
$\VR$ is accurately given by
\begin{equation}
  \label{eqn:LMFeqn}
  \VR(\vect{r}) = \mathcal{V}(\vect{r}) + \frac{1}{4\pi\epsilon_0}\int \rho_R^{q}(\vect{r}^\prime;[\VR]) v_1\left(\left|\vect{r}^\prime - \vect{r}\right|\right) \, d\vect{r}^\prime + C.
\end{equation}
The notation $\rho^q_R(\vect{r};[\VR])$ simply indicates that we now
use the charge density profile of the mimic system
(indicated by $R$) in the presence of the restructured potential $\VR$.
The charge density $\rho^q_R$ encompasses the effect of $v_0(r)$ and
$\VR$ as well as all other non-electrostatic interactions like the
hydrophobic walls and the LJ cores in SPC/E water.  We use the constant
$C$ to set the zero of our potential.  In the systems examined in this
paper, this formula may be rewritten as a one-dimensional equation as
detailed in
Ref.~\cite{ChenWeeks.2006.Local-molecular-field-theory-for-effective}.

$\VR$ can be understood as a mean-field electrostatic potential
representing the long-ranged forces that may be reasonably averaged
over.  Alternately, harnessing the convolution definition of $v_1(r)$
in Eq.~\ref{eqn:v1def}, we may rewrite the LMF equation as the
convolution of a smoothed charge density with the full Coulomb
potential. Thus, $\VR$ exactly satisfies a modified Poisson's equation
\begin{equation}
  \label{eqn:LMFPoisson}
  \laplacian \left\{\VR(\vect{r}) - \mathcal{V}(\vect{r}) \right\} = - \frac{1}{\epsilon_0} \rho_R^{q_\sigma}(\vect{r};[\VR]),
\end{equation}
defined with a Gaussian-smoothed charge density
\begin{equation}
  \label{eqn:SmoothChg}
  \rho^{q_\sigma}(\vect{r}) = \frac{1}{\pi^{3/2}\sigma^3}\int \rho^q(\vect{r}^\prime) e^{-\left|\vect{r}-\vect{r}^\prime\right|^2/\sigma^2} \, d\vect{r}^\prime.
\end{equation}
This again emphasizes that \emph{all} charge interactions in the
system must be identically truncated for this modified electrostatics
to be valid.

The form of the short-ranged potential $v_0(r)$ might suggest an
analogy to Ewald summation.  However, LMF theory focuses on the
short-ranged mimic system and how best to use and correct its
properties based on results from equilibrium theory, while Ewald
summation focuses on the full system with exact periodic boundary
conditions.  As a result, $\VR$ is a static external potential based
on equilibrium properties that is determined self-consistently as
described in the Methods below, whereas Ewald summation evaluates
fluctuating forces arising from both the short- and long-ranged
components of the Coulomb interactions at each time step.
  
\section{Classical Failure of GT Water Explained}
While we believe that the above failure of GT water alone
in an electric field has not been noted before, problems with
electrostatic properties of confined truncated water in zero field are well
known~\cite{FellerPastorRojnuckari.1996.Effect-of-Electrostatic-Force-Truncation-on-Interfacial}.
Shown in Fig.~3 is the classic example of the failure of GT
water in nonuniform situations -- the incorrect induced potential
profile of water molecules.  SPC/E water is again confined between the
hydrophobic walls, now with no applied electric field.  As
well-established by Rossky and
coworkers~\cite{LeeMcCammonRossky.1984.The-Structure-of-Liquid-Water-at-an-Extended-Hydrophobic},
water molecules next to the wall tend to point one hydroxyl group into
the wall so that one hydrogen bond per molecule on average is broken
rather than two.  This induces a dipole layer near each hydrophobic
surface with the dipole pointing toward the walls.  This
atomically-detailed dipole layer leads to a semi-discrete jump at the
surface in the electrostatic potential defined as
\begin{equation}
  \label{eqn:elecpotential}
  \Phi_{\text{pol}}(z) = -\frac{1}{\epsilon_0}\int_{-L/2}^{z} dz^\prime \int_{-L/2}^{z^\prime} dz^{\prime\prime} \rho^{q}(z^{\prime\prime}).
\end{equation}
This potential $\Phi_{\text{pol}}$ should plateau in the central bulk
region; however, the use of GT water fails dramatically in this
respect.  Due to the initial overorientation of water molecules at the
surface, driven by the maintenance of hydrogen bonds in GT water, the
polarization potential \emph{never} plateaus.  This failure is well
documented~\cite{FellerPastorRojnuckari.1996.Effect-of-Electrostatic-Force-Truncation-on-Interfacial,Spohr.1997.Effect-of-Electrostatic-Boundary-Conditions-and-System}
and is attributable to the neglect of net-additive long-ranged forces
due to the formation of dipole layers at the surface. As also shown in
Fig.~3, LMF theory correctly includes those neglected forces with
tremendous success.

\begin{figure}
  \centering
  \includegraphics{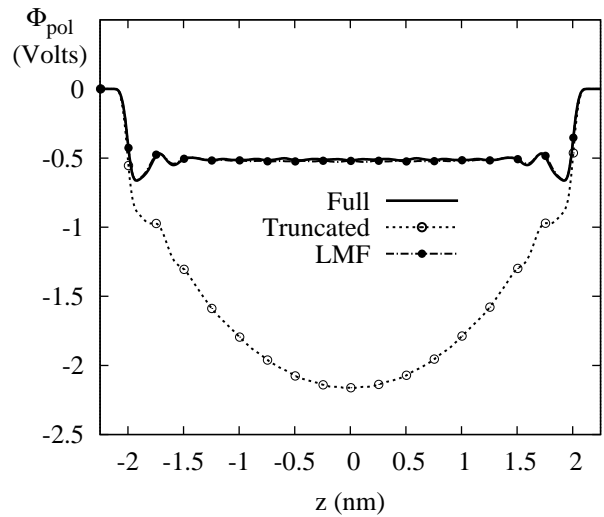}
  \caption{Polarization potential profile,$\Phi_{\text{pol}}(z)$, between hydrophobic walls with no applied field.  GT water does not display the expected plateau in the bulk region.  Treatment of the system with LMF theory yields strong agreement with full treatment of the electrostatics.}
\end{figure}

The form of $\VR(z)$ shown in Fig.~4 highlights its physical effect on
GT water molecules.  Without $\VR$, the truncated molecules overorient
near the wall.  The restructured electrostatic potential $\VR$ from LMF
theory reflects the net long-ranged forces due to the ordering of
water near the surfaces by applying a reorienting torque on surface
water molecules.  Since $\VR$ is defined via a self-consistent
equation, it represents a statistical mechanical balance between the
favorability of maintaining local hydrogen bonds and the electrostatic penalty for
creating an overly severe dipole layer.  The potential $\VR$ varies smoothly
with $z$ because it is determined by averaging over the
slowly-varying $v_1(r)$ in Eq.~\ref{eqn:LMFeqn}.
This suggests that a simple theory yielding this general form may be possible.

\begin{figure}
  \centering
  \includegraphics{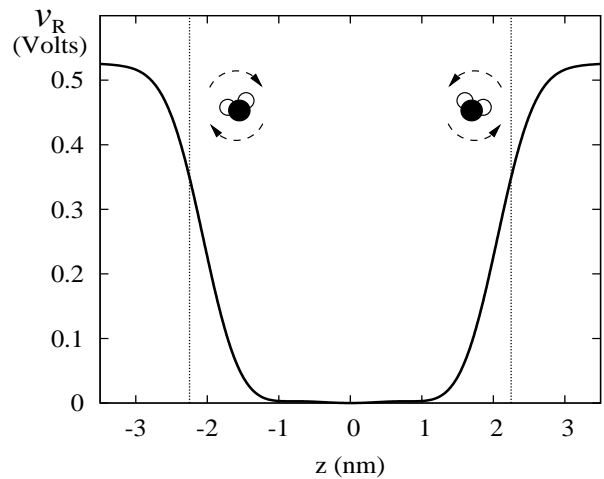}
  \caption{Form of $\VR(z)$ for water confined between hydrophobic walls with no applied electric field. 
The self-consistent potential applies a reorienting torque on water molecules near the surface,
mimicking the effect of the long-ranged dipole layers.}
\end{figure}

While the physics of this failure was convincingly established by
Spohr~\cite{Spohr.1997.Effect-of-Electrostatic-Boundary-Conditions-and-System},
its structural roots in the charge density profile remained
mysterious.  Brooks and coworkers observed an overorientation of
molecular dipoles near the walls in spherically truncated
water~\cite{FellerPastorRojnuckari.1996.Effect-of-Electrostatic-Force-Truncation-on-Interfacial}.
But as shown in Fig.~5\emph{a}, the charge density profile for GT
water tracks quite closely with the results from full Ewald summation
techniques for SPC/E water.  A dipole layer near the wall seems quite
clear, and there are only slight deviations in the peaks next to the
walls.  However, the polarization potential $\Phi_{\text{pol}}$ belies
the apparent accuracy of the atomic-level charge density profiles
for GT water. The
polarization potential may be fit accurately by a quadratic function
in the central region between the walls.  This fit suggests the
existence of a constant residual charge density of approximately
$-0.04$~e$_0$/nm$^3$ throughout the ``bulk'' region neutralizing the
overaccumulation of positive charge next to each wall.

\begin{figure}
  \centering
  \includegraphics{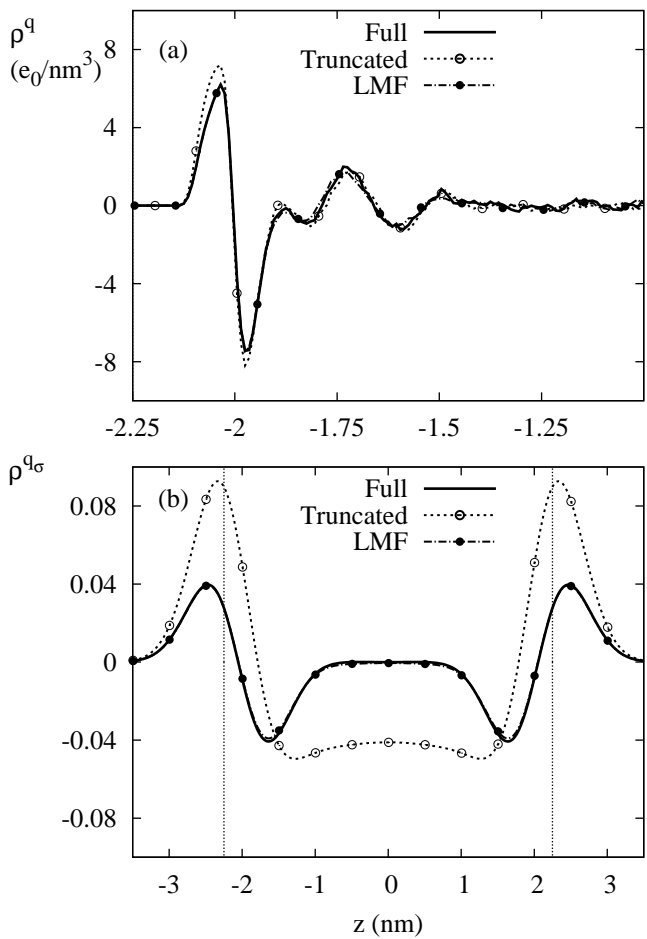}
  \caption{Charge density profiles for water confined between hydrophobic walls with no applied field.  (a)~The atomic level charge density $\rho^q$ for $z<1.0$~nm shows only slight discrepancies between GT water and either full or LMF-treated water.  This density profile is symmetric about the origin. (b)~The gaussian-smoothed charge density $\rho^{q_\sigma}$ shows the failure of GT water to capture the true electrostatic dipole layer much more clearly.  In either instance, the LMF-treated water is in very good agreement with a full treatment of electrostatics.}
\end{figure}

LMF theory allows us to highlight exactly that relevant structural
problem.  The writing of the LMF equation as a modified Poisson's
equation in Eq.~\ref{eqn:LMFPoisson} suggests that the charge density
that best reflects the long-ranged electrostatics is not $\rho^q$, but
rather the Gaussian-smoothed charge density $\rho^{q_\sigma}$.  The
smoothed charge density $\rho^{q_\sigma}$ in Fig.~5\emph{b} clearly
shows the dipolar layer at each wall for the water confined with no
electric field.  Beyond that, GT water alone very obviously spreads
the dipole layer into the bulk.  The residual negative charge density
is approximately the value suggested by the form of
$\Phi_{\text{pol}}$ for GT water. Since the smoothed charge density is
two orders of magnitude smaller than the atomic level charge density,
this underlying electrostatics is buried alongside the slight
simulation noise in $\rho^q$.  On Gaussian-smoothing, the random
simulation noise in $\rho^q$ cancels, uncovering the smaller but
coherent electrostatic features.  In a sense, the strong charge
layering over 0.2 nm displayed in Fig.~5\emph{a} is \emph{not} the
long-ranged dipole near the wall.  The true dipole layer is due to
average behavior over a few molecular layers.  Classical electrostatic
textbooks often posit some sort of smoothing over such local
order~\cite{CorsonLorrain.1962.Introduction-to-Electromagnetic-Fields-and-Waves}.

This smoothed charge density allows for a more direct analogy between
molecular simulations and classical electrostatics.  The LMF approach
naturally distinguishes the long-ranged charge-charge interactions
representing overall dipolar ordering from the short-ranged
charge-charge interactions in simulations that capture
hydrogen-bonding and short-ranged polar attraction.  Given that any
$\sigma$ greater than some $\sigma_{\text{min}}$ will describe
relevant short-ranged interactions like hydrogen bonding and can be
used in simulations, too much significance should not be assigned to
the particular chosen $\sigma$ value or the exact magnitudes of the
resulting $\rho^{q_\sigma}$.  However we strengthen a previous
suggestion~\cite{HummerPrattGarcia.1997.Electrostatic-potentials-and-free-energies-of-solvation}
that charge density profiles reveal electrostatic effects in a
less-biased fashion than molecule-based profiles: perhaps a smoothed
profile is even less biased.

\section{LMF Theory Is More Broadly Applicable}
We now illustrate the application of LMF theory to moderately more realistic
molecular surfaces by replacing the hydrophobic walls with empirical
Pt(111) surfaces that order the water molecules at the surface,
attracting the oxygen atoms to localized binding sites as shown by
Berkowitz and
coworkers~\cite{RaghavanFosterMotakabbir.1991.Structure-and-Dynamics-of-Water-at-the-Pt111-Interface:}.
This surface is meant to represent specific interactions and detailed
ordering of water molecules at an atomically corrugated surface, thus
mimicking a hydrophilic interaction.  The particular wall potential we
use does not have explicitly fixed wall charges, but LMF theory can
readily handle these as well~\cite{RodgersWeeks.2008.Local-molecular-field-theory-for-the-treatment}.

Since this surface induces a charge density that is explicitly a
function of $x$, $y$, and $z$, one might reasonably conclude that we
now must solve the full three-dimensional LMF equation given in
Eq.~\ref{eqn:LMFeqn} to obtain a $\VR$ that is a function of
$\vect{r}$.  However, the charge smoothing inherent in the LMF
equation justifies a much simpler treatment of $\VR$.  The lateral
variations in charge density are washed out when smoothed with the
Gaussian of width $\sigma$, leaving only a slowly-varying
$\rho^{q_{\sigma}}(z)$ in the Poisson-like formulation of the LMF
equation given in Eq.~\ref{eqn:LMFPoisson}.  This $\rho^{q_\sigma}(z)$
again reflects the formation of dipole layers near each wall.  Thus,
solving a one-dimensional LMF equation yields remarkably accurate
results for the polarization potential profile $\Phi_{\text{pol}}(z)$,
as shown in Fig.~6.  Such findings give us confidence that the LMF
treatment of interesting and more complex simulations such as
water-membrane interfaces and liquid-liquid interfaces will prove to
be just as simple.

\begin{figure}
  \centering
  \includegraphics{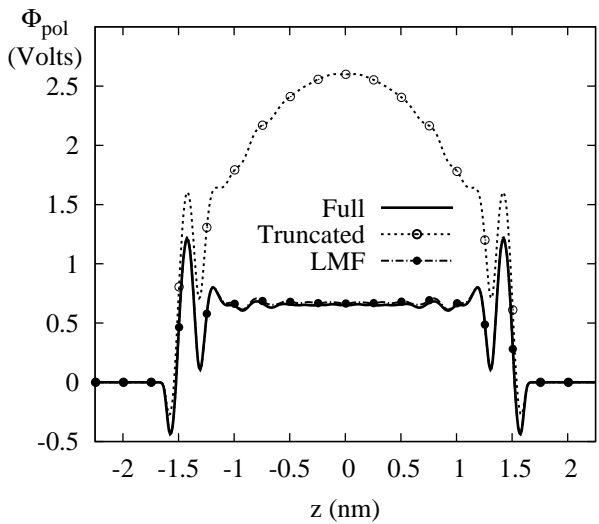}
  \caption{Polarization potential profile,$\Phi_{\text{pol}}(z)$, between corrugated walls with no applied field.  GT water again does not display the expected plateau in the bulk region.  Treatment of the system with LMF theory again yields strong agreement with full treatment of the electrostatics.}
\end{figure}

\section{Conclusions and Outlook}
Here we have demonstrated the utility of the LMF approach both in
simulating a molecular system in a slab geometry and also in analyzing
the structural properties of this system.  The inclusion of a restructured
electrostatic potential $\VR$ allows for remarkably accurate results
in this slab geometry.  The spectacular failure of the spherical
truncation of $1/r$ interactions to describe even the density profile
of water in an applied field is corrected through the use of LMF
theory.  Earlier known failures of truncated water models to describe
the electrostatic potential for confined water in zero field are also
corrected by LMF theory regardless of the specific details of the
confining surfaces.  Furthermore, the splitting of $1/r$ into
short-ranged and long-ranged components provides a natural way to
disentangle the impact of local hydrogen-bonding and of long-ranged
dipolar interactions in various aqueous systems.  In light of LMF
theory's success in this slab geometry, we certainly believe that it
could be of use in other physically relevant slab-geometry simulations
such as biological membranes and liquid-liquid interfaces.  Perhaps it
also could help resolve a recent dispute over whether or not depletion
of water between capacitor plates occurs in the simulation of an open
system~\cite{BratkoDaubLeung.2007.Effect-of-field-direction-on-electrowetting-in-a-nanopore,VaitheeswaYinRasaiah.2005.Water-between-plates-in-the-presence-of-an-electric,EnglandParkPande.2008.Theory-for-an-order-driven-disruption-of-the-liquid-state}.
Furthermore, given previous work treating primitive models of ionic solutions with LMF
theory~\cite{ChenKaurWeeks.2004.Connecting-systems-with-short-and-long,ChenWeeks.2006.Local-molecular-field-theory-for-effective,DenesyukWeeks.2008.A-new-approach-for-efficient-simulation-of-Coulomb-interactions},
the study of aqueous ionic solutions is another natural step
forward. Truncating all $1/r$ interactions via $v_0(r)$ with the same
$\sigma$ would lead to a natural mimic system for such ionic
solutions.

Examination of either
$\VR$ or the smoothed charge density $\rho^{q_\sigma}$ prescribed by
LMF theory also clarifies the underlying electrostatics and the failure of
spherical truncations. The Gaussian-smoothed perspective for long-ranged
electrostatics greatly elucidates the perceived contradiction between
seemingly good charge density profiles but very poor polarization
potential profiles for water confined between walls in zero field.
Analysis of long-ranged electrostatic properties using the smoothed
charge density seems more plausible because the point charges in
classical site-site models approximate the real quantum-mechanical
electron cloud, and longer-ranged electrostatic effects should be
relatively insensitive to the exact placement and magnitude of point charges in
models with the same dipole moment. Thus in GT water, the short-ranged cores are altered to reflect
both hydrogen-bonding and polar
attractions represented by $v_0(r)$ as well as the LJ cores.  Simulation using GT water
without a self-consistent $\VR$ will only be reasonable if the
resulting $\rho^{q_\sigma}$ reflects the underlying electrostatics.
Since $\VR$ is the solution to a modified Poisson's equation, we
should not expect this to be generally true.

Overall the use of LMF theory for electrostatics in simulations seems
very promising, but certainly this approach needs more investigation.
Upcoming articles will detail the derivation of the simple LMF
electrostatic potential given in Eq.~\ref{eqn:LMFeqn} for charge
mixtures~\cite{RodgersWeeks.2008.Local-molecular-field-theory-for-the-treatment}
and for site-site molecular models~\cite{HuRodgersWeeks..} as well as
simple analytical thermodynamic corrections for these mimic
systems~\cite{RodgersWeeks..}.  Further investigation into optimizing
the solution of the LMF equation is called for, and the slowly-varying
nature of the smoothed charge density suggests that simple
approximations could prove quite
useful~\cite{DenesyukWeeks.2008.A-new-approach-for-efficient-simulation-of-Coulomb-interactions}.
At present, the LMF mapping is for equilibrium properties only.
However current work has shown that LMF theory used in tandem with
with a Born-Oppenheimer-like approximation can describe the slow
dynamics and the associated free energy landscape of charged polymers
folding in the presence of mobile salt ions~\cite{DenesyukWeeks..}.
In general, however, a full extension of these ideas to dynamics
remains a challenging open question. Its resolution may clarify the
relationship between the fluctuations of these short-ranged systems
and those in experimental systems.

While all nonuniform systems examined herein have a slab geometry, LMF
theory is valid for general nonuniformity.  As such, we view the work
presented in this paper as a demonstration of the utility of LMF
theory to treat and to understand electrostatic interactions in
molecular systems far more complex than those studied here.

\section{Methods}
For this research, we have used the {\sc dlpoly2.16} molecular
dynamics
package~\cite{SmithYongRodger.2002.DLPOLY:-Application-to-molecular-simulation}
modified to include the Lennard-Jones wall
potential~\cite{LeeMcCammonRossky.1984.The-Structure-of-Liquid-Water-at-an-Extended-Hydrophobic},
the Pt(111) wall
potential~\cite{RaghavanFosterMotakabbir.1991.Structure-and-Dynamics-of-Water-at-the-Pt111-Interface:},
the LMF theory $v_0(r)$ potential and the $\VR(z)$ mean field
potential, as well as the three-dimensional corrected Ewald sum for
slab systems
(EW3DC)~\cite{YehBerkowitz.1999.Ewald-summation-for-systems-with-slab}.
SPC/E
water~\cite{BerendsenGrigeraStraatsma.1987.The-missing-term-in-effective-pair-potentials}
was simulated at constant volume and temperature using the Berendsen
thermostat~\cite{BerendsenPostmaGunsteren.1984.Molecular-dynamics-with-coupling-to-an-external-bath}
with a timestep of 1~fs.  For the bulk water pair correlation
functions, we simulated 1728 SPC/E molecules at bulk density for
1.5~ns after 500~ps of equilibration at a temperature of 300~K. For
each nonuniform system and electrostatic technique we also have
carried out at least 500~ps of equilibration and subsequently 1.5~ns
of simulation at 298~K.  The lateral simulation box size for the
smooth Lennard-Jones walls was 2.772~nm~$\times$~2.772~nm with 1024
SPC/E molecules, and the walls were placed at $\pm 2.25$~nm in order
to obtain the bulk density of water in the central region.  When the
electric field of 10.0~V/nm was applied, the walls were moved to
2.153~nm to maintain the bulk density of water.  (Therefore effects of
electrostriction were not accounted for.)  The Pt(111) wall area was
3.05~nm~$\times$~2.88~nm with 1054 SPC/E molecules, and the $z_c$ for
the walls (as defined in
Ref.~\cite{RaghavanFosterMotakabbir.1991.Structure-and-Dynamics-of-Water-at-the-Pt111-Interface:})
were placed at $\pm 1.56$~nm.  In both cases the simulation cell
spanned 14.0~nm in the $z$-direction, though this was only crucial for
the Ewald summations to maintain sufficient vacuum space between
periodically replicated slabs.

The three electrostatic techniques used for the slab geometry were
Ewald
summation~\cite{Ewald.1921.Evaluation-of-optical-and-electrostatic-lattice-potentials},
minimum image simulations using a spherically-truncated $v_0(r)$, and
minimum image simulations using a spherically-truncated $v_0(r)$ and
$\VR(z)$.  The Ewald technique used was the corrected
three-dimensional Ewald scheme of Yeh and
Berkowitz~\cite{YehBerkowitz.1999.Ewald-summation-for-systems-with-slab}.
We used an $\alpha$ of 3.40~nm$^{-1}$ and $k$-vectors spanning
(12,12,60) for LJ walls and (13,12,60) for Pt(111) walls.  For
$v_0(r)$ we used a $\sigma$ of 0.60~nm for slab simulations and
0.40~nm for the bulk simulations.  For bulk, using a larger $\sigma$
of 0.60~nm produced equivalent results, though the $\delta g$ will be
slightly different.

The LMF equation must be solved self-consistently so that the $\rho^q$
determined via simulation using $\VR$ yields the same LMF equation
prediction of $\VR$.  Self-consistently solving the LMF equation in
previous simulation
work~\cite{RodgersKaurChen.2006.Attraction-between-like-charged-walls:-Short-ranged}
proved relatively straightforward with a linear mixing parameter
$\lambda$ as $\VR^{i+1} = \lambda \cdot \VR^{\text{LMF}} +
(1-\lambda)\cdot \VR^{i}$ where $\VR^{\text{LMF}}$ is the restructured
potential determined by inserting the $\rho^q$ due to $\VR^{i-1}$ into
Eq.~\ref{eqn:LMFeqn}.  However, for these systems, the solution proved
more challenging due to the need to distinguish between inherent
equilibrium fluctuations in molecular orientations in the short-ranged
system and the induced response of the short-ranged system to a
variation in the small and slowly-varying $\VR$.

We may still solve the LMF equation fairly quickly, given a reasonable
initial $\VR^0$, by carrying out $N$ simultaneous short-time
simulations with distinct starting points during the evolution to a
self-consistent solution.  The averaged $\rho^q$ of the $N$ systems
simulated in $\VR^i$ at the $i^{th}$ solution step better revealed the
response of the system to the change in $\VR$ from $\VR^{i-1}$ to
$\VR^i$.  These short-time simulations each ran for 25~ps of
equilibration and 50~ps of accumulation with a larger $\Delta$t of
2.5~fs.  Then the $\rho^q(z)$ resulting from each of the $N$
short-time simulations are averaged together and inserted in the LMF
equation to yield $\VR^{\text{LMF}}$.  Just as for the simpler
solution schemes, we then used linear mixing with $\lambda$ ranging
from 0.05 to 0.10 to predict the subsequent $\VR^{i+1}$.  For
$\sigma=0.60$~nm, $\VR$ converged within 3 steps of 10 parallel
simulations with no applied $E_0$ and within 10 steps of 10 parallel
simulations with an applied field of 10.0~V/nm.  However, this process
was employed simply to demonstrate the utility of a self-consistent
LMF treatment of electrostatics in molecular simulations.  An
optimized approach will likely appear quite different.  Currently,
using the interpretation of the LMF equation given in
Eq.~\ref{eqn:LMFPoisson}, we are exploring options such as fast
Poisson solvers and Car-Parinello techniques in solving the equation
adaptively during simulation.  Such solutions for $\VR$ would still be
static in nature once self-consistency was achieved.

\begin{acknowledgments}
  This work was supported by NSF grant \# CHE-0517818.  JMR
  acknowledges the support of the University of Maryland Chemical
  Physics fellowship.  The computations were supported in part by the
  University of Maryland and in part by the National Science
  Foundation through TeraGrid resources provided by NCSA and TACC.
  Our initial interest in water in an electric field was stimulated by
  a provocative discussion with Max Berkowitz.  We also thank Michael
  Fisher for very helpful comments on the structure of this paper.
\end{acknowledgments}











\end{document}